\pdfoutput=1
\documentclass[onecollarge,natbib]{svjour2}
\bibpunct{[}{]}{;}{n}{}{,} 
\smartqed  
\usepackage{graphicx,amsmath}
\journalname{}
\def\be{\begin{eqnarray} &&}

\def\nonu{\nonumber \\ &&}
\def\ee{\end{eqnarray}}
 \usepackage{wrapfig}
 \usepackage[normalem]{ulem}
 \usepackage{graphicx,color}
 \usepackage{enumerate}

     \font\tenbifull=cmmib10 scaled 1200 
     \font\tenbimed=cmmib9
     \font\tenbismall=cmmib7
\mathchardef\bbkappa="7114
\mathchardef\bbrho="711A
\mathchardef\bbsigma="711B
\mathchardef\bbtau="711C
\mathchardef\bbvarrho="7125
\mathchardef\bbvarsigma="7126
\mathchardef\bbxi="7118

\def\be{\begin{eqnarray} &&}

\def\nonu{\nonumber \\ &&}
\def\ee{\end{eqnarray}}
\newcommand\la{\langle}
\newcommand\ra{\rangle}

\makeindex

\def\beq{\begin{equation}}
\def\eeq{\end{equation}}
\usepackage{amsbsy,amsmath}

\newcommand{\blf}[1]{\bf  {\tilde #1}}
\newcommand{\bq}{\begin{eqnarray}}
\newcommand{\eq}{\end{eqnarray}}

\def\sumint{\int \! \!\ \! \! \! \! \!\ \! \! \!\! \!\sum}

\newcommand {\vece}[1]{\overset{_\rightarrow}{#1}}
\begin{document}

\title{  Towards an improved 
description
of  SiDIS by a polarized $^3$He target
}


\author{Alessio Del Dotto  \and Leonid Kaptari \and Emanuele Pace 
\and    Giovanni Salm\`e \and  Sergio Scopetta  }


\institute{Alessio Del Dotto \at
              Universit\`a di Roma Tre and INFN, Roma, Italy \and 
           Leonid Kaptari \at
           Bogoliubov LTP, JINR, Dubna, Russia
           \and
           Emanuele Pace \at
              Universit\`a di Roma ``Tor Vergata'' and INFN, Roma, Italy
               \and
          Giovanni Salm\`e \at
             INFN, Sezione di Roma, Italy
               \and
          Sergio Scopetta  \at
             Universit\`a di Perugia and
          INFN, Sezione di Perugia, Italy
}
\date{}

\maketitle

\begin{abstract}
The possibility of  improving the description of the semi-inclusive deep inelastic
electron scattering off
 polarized $^3$He, that provides information on   the neutron single 
spin asymmetries, is illustrated. In particular,
 the  analysis
 at finite  momentum transfers in a Poincar\'e
covariant framework is outlined and a generalized eikonal approach to include final
state interaction is presented.
\keywords{neutron  spin asymmetries \and Light-front dynamics 
\and final-state interaction \and polarized $^3$He target
}
\end{abstract}

\section{Introduction}
\label{intro}

As it is well known, the quark { helicity} takes into account at most one third of the
 proton spin. This motivates the great effort on both experimental and theoretical
sides  to accurately determine the contributions from the
quark orbital angular momentum ($L_q$) and from the gluons. 
 Analogous efforts are carried out  to study the neutron. In view of this,
the investigations of reactions like the electron 
  semi-inclusive deep inelastic  scattering  (SiDIS) by a 
 polarized $^3$He target play
a very relevant role \cite{He3exp}. Such processes yield information on 
 the so-called transverse-momentum distributions (TMD) of a polarized 
quark
inside a polarized neutron, such that  an unprecedented amount of details 
on the quark dynamics inside the neutron can be achieved. Our aim
\cite{todo} is to extend a  previous analysis of SiDIS
by polarized $^3$He target \cite{Sco}, that 
 addressed  the  extraction    of
 the neutron single-spin asymmetries (SSAs), within  a plane wave impulse 
 approximation (PWIA) framework and in the Bjorken limit.  One can improve
  such an
approach by: i) dealing with the  
relativistic effects through
a  Poincar\'e covariant description of the nuclear dynamics (see,
e.g.,   \cite{Dotto,Pac})  and ii) taking   into account 
  the final state interaction (FSI) by adopting \cite{our} the so-called 
generalized eikonal approach (GEA) (see, e.g. \cite{ourlast}).

\section{The polarized $^3$He nucleus  as an effective neutron target}
\label{sec:1}
A polarized $^3$He nucleus  is an ideal target to study the neutron, since at 
a 90\% level it is equivalent to a polarized neutron. For disentangling the nucleon
structure from the  
dynamical nuclear effects, one can 
adopt  an approach based on the spin-dependent spectral function of 
$^3$He, ${\rm P}_{\sigma,\sigma^\prime} (\vec p, E)$, (see, e.g.  \cite{Ciofi1})
that yields
the probability distribution to find a nucleon with given missing energy,
three-momentum and polarization inside the nucleus.
By using  this formalism,  one can safely extract \cite{Ciofi2} the 
neutron longitudinal asymmetry, $A_n $,
from the corresponding $^3$He  observable, $A_3^{exp}$, 
obtained from   the reaction
 $^3\vece{{\rm He}}( \vece{e},e')X$ in DIS regime, i.e.   
\be
  {{A_n }}\simeq  \left 
( {A^{exp}_3} - 2 
{p_p} f_p
{{A^{exp}_p}} \right )/(p_n f_n )\label{dis}\ee
with  $p_{n(p)}$  the
neutron (proton) effective polarization inside the polarized $^3$He, and
$f_{n(p)}$,  the dilution factor. Realistic values
of $p_n$ and $p_p$ are  
$ {{p_p}} = -0.023  $, 
${{p_n}}= 0.878 $  (see, e.g., \cite{Ciofi2,Sco}).
  In \cite{Sco}, an analogous extraction was applied to the SSA of a
  transversely polarized $^3\vece{{\rm He}}$ target, obtained 
  from
the  process $^3\vece{{\rm He}}( e,e'\pi)X$, in order to obtain the 
SSA of a transversely polarized neutron. 
 In PWIA and adopting  the Bjorken limit, the SSAs of  $^3\vece{\rm { He}}$
are a  convolution of  
$ {\rm P}_{\sigma,\sigma^\prime} (\vec p, E)$, and the nucleon SSAs, that in turn
are convolutions of suitable 
  TMDs
 and  fragmentation functions, phenomenologically describing  the hadronization
 of the hit quark. 
To improve the previous description, taking into account 
 relativistic effects in the 
actual experimental kinematics and developing a Poincar\'e covariant
framework for analyzing the SSA of $^3$He, one can  adopt  the Light-Front (LF)  
Relativistic Hamiltonian Dynamics (RHD), combined with
 the Bakamjian-Thomas 
construction of the Poincar\'e generators. 
Then, one gets the following expression
for the 
$^3$He  hadronic tensor \cite{Dotto,Pac} (for details see \cite{todo})
\be
{{ 
{\cal W}^{\mu\nu}(Q^2,x_B,z,\tau, \hat{\bf h},S_{He})}} 
 \propto
 \sum_{\sigma,\sigma'}\sum_{\tau} 
 \left.\sumint \right._{\epsilon^{min}_S}^{\epsilon^{max}_S}{~d
\epsilon_{S} }
\int_{M^2_N}^{(M_X-M_S)^2} dM^2_f \quad \quad
\nonu
 \times 
\int_{\xi_{l}}^{\xi_{u}} {d\xi\over \xi^2 (1-\xi)(2\pi)^3}
\int_{P^{m}_\perp}^{P^{M}_\perp}{d
P_\perp\over sin\theta   }~ (P^++q^+- h^+)
~ 
{{
w^{\mu\nu}_{\sigma\sigma'}\left(\tau,{\blf q},{\blf h},{\blf P}\right)
}}
{{
{\cal P}^{\tau}_{\sigma'\sigma}(\tilde{\bf k},\epsilon_S,S_{He})
}}
\ee
where $ \tilde{\bf v} = \{v^+=v^0+v^3, {\bf v_{\perp} } \}$,
$ w^{\mu\nu}_{\sigma\sigma'}
\left(\tau,{\blf q},{\blf h},{\blf P}\right)$
is the hadronic nucleon  tensor and
$
{\cal P}^{\tau}_{\sigma\sigma'}(\tilde{\bf k},\epsilon_S,S_{He})
$
 the {{LF}} spin-dependent spectral function,  related to the instant-form
  spectral function, $ {\cal S}^{\tau}_{\sigma'_1\sigma_1}({\bf
 k},\epsilon_{S},S_{He})$ through
 the unitary Melosh Rotations, 
$
{{D^{{1 \over 2}} [{\cal R}_M ({\blf k})]}}$, as follows
\be
{ {
  {\cal P}^{\tau}_{\sigma'\sigma}({\blf k},\epsilon_{S},S_{He})
}}
\propto
~\sum_{\sigma_1 \sigma'_1} 
{{D^{{1 \over 2}} [{\cal R}_M^\dagger ({\blf
k})]_{\sigma'\sigma'_1}}}~
{{
{\cal S}^{\tau}_{\sigma'_1\sigma_1}({\bf k},\epsilon_{S},S_{He})
}} ~
{{D^{{1 \over 2}} [{\cal R}_M ({\blf k})]_{\sigma_1\sigma}}}
\label{sfrhd}
\ee
In our approach, 
$
{\cal S}^{\tau}_{\sigma'_1\sigma_1}
$
can be approximated by  ${\rm P}^{\tau}_{\sigma'_1\sigma_1}$, obtained within 
a non relativistic framework,
since the constraints imposed by the  Poincar\'e covariance can be 
fully
satisfied with our assumptions, namely the LF RHD framework completed by the 
  Bakamjian-Thomas construction \cite{todo}.

\section{Beyond PWIA: the generalized eikonal approximation }
The second ingredient to be added  is 
GEA (see,
e.g., \cite{ourlast} and references therein quoted), devised
for taking into account the FSI effects. In particular, FSI effects  to be considered 
 are   due to the propagation   of the 
   debris, formed after  the $\gamma^*$ absorption by a target quark,  and the
   subsequent
   hadronization,  both of them influenced by the  presence of a
    fully-interacting $(A-1)$ spectator system. Clearly,  such  FSI effects 
    represent  a very complicated many-body problem
and  their evaluation from first principles is a hard challenge. Therefore  it is
necessary 
to  develop model approaches for evaluating their impact on the extraction of
neutron SSAs.
The approximation based on GEA has been recently
applied for describing the spectator SiDIS by a polarized $^3$He target
\cite{our}. The key quantities are 
the intrinsic overlaps defined as follows
\be\!\!\!\!\!\!\!\!
{\cal O}^{\hat {\bf S}_A(FSI)}_{\lambda\lambda'}({\bf  p}_N,E) = ~
\sum \! \!\! \!\! \!\! \!\int{~d\epsilon^*_{A-1}}~\rho\left(
\epsilon^*_{A-1}\right)~
 \la  \hat{S}_{Gl} \{ \Phi_{\epsilon^*_{A-1}},\lambda,{\bf p}_N\}|  S_A, \Phi_A\ra
 \la   S_A, \Phi_A| \hat{S}_{Gl}\{ \Phi_{\epsilon^*_{A-1}},\lambda',{\bf p}_N\} \ra
 ~\times \nonu ~~~~~~~~~
 \delta\left( { E+ M_A-m_N-M^*_{A-1}-T_{A-1}}\right)
\label{overlaps}
\ee
where i) $E$ is 
the usual missing energy $E=\epsilon^*_{A-1}+B_A$, with 
$\epsilon^*_{A-1}$ ($\rho\left(
\epsilon^*_{A-1}\right)$)  the energy (state density) of the spectator 
system and  $B_A$ the binding
energy of the target nucleus, ii) ${\bf  p}_N$ the three-momentum of the struck nucleon
and iii) 
$\hat S_{Gl}(1,2,3)$ represents  the debris-nucleon eikonal
scattering S-matrix, that depends  upon the relative coordinates only, and  it is
given by 
 $
\hat S_{Gl} ({\bf r}_1,{\bf r}_2,{\bf r}_3)=
\prod_{i=2,3}\bigl[1-\theta(z_i-z_1)\Gamma({\bf b}_1-{\bf b}_i,{ z}_1-{z}_i)
\bigr]
$,
where  ${\bf b}_i (z_i)$
is the perpendicular (parallel) component of
${\bf r}_{i}$ (remind that ${\bf r}_1+{\bf r}_2+{\bf r}_3=0$), with
respect
to the direction of the propagation of the debris ${\bf p}_X$.
In the DIS limit
 ${\bf p}_X \simeq{\bf q} $  (${\bf q}$ is the three-momentum transfer) and the
eikonal S-matrix
is defined with respect to $\bf q$. 
The profile function, $\Gamma$, is \cite{ourlast}
\be
\Gamma({{\bf b}_{1i}},z_{1i})\, =\,\frac{(1-i\,\alpha)\,\,
\sigma_{eff}(z_{1i})} {4\,\pi\,b_0^2}\,\exp \left[-\frac{{\bf
b}_{1i}^{2}}{2\,b_0^2}\right]~~~,
 \label{eikonal}
\ee
where  ${ z}_{1i} ={ z}_{1}-{
z}_{i}$,   ${\bf b}_{1i}={\bf b}_{1}-{\bf b}_{i}$, and
$\sigma_{eff}(z_{1i})=\sigma^{NN}_{tot}(z_{1i})+\sigma^{\pi N}_{tot}(z_{1i})~
N^\pi_{eff}$, 
with $N^\pi_{eff}$ the
effective number of pions which are produced.
Indeed, the effective cross sections, $\sigma_{eff}(z_{1i})$
 depends also on the total energy of the debris, 
 $W^2\equiv p_X^2={ (p _N +q)^2}$,  but such a dependence is weak,
 if the energy is not too large and the
hadronization process develops inside the nuclear environment.
Hence, 
the effective cross section can be approximated as
$\sigma_{eff}(z_{1i},x_{Bj},Q^2)\sim \sigma_{eff}(z_{1i})$
\cite{ourlast,ckk}.
\begin{figure}          
\parbox{8cm}{\includegraphics[width=8cm]{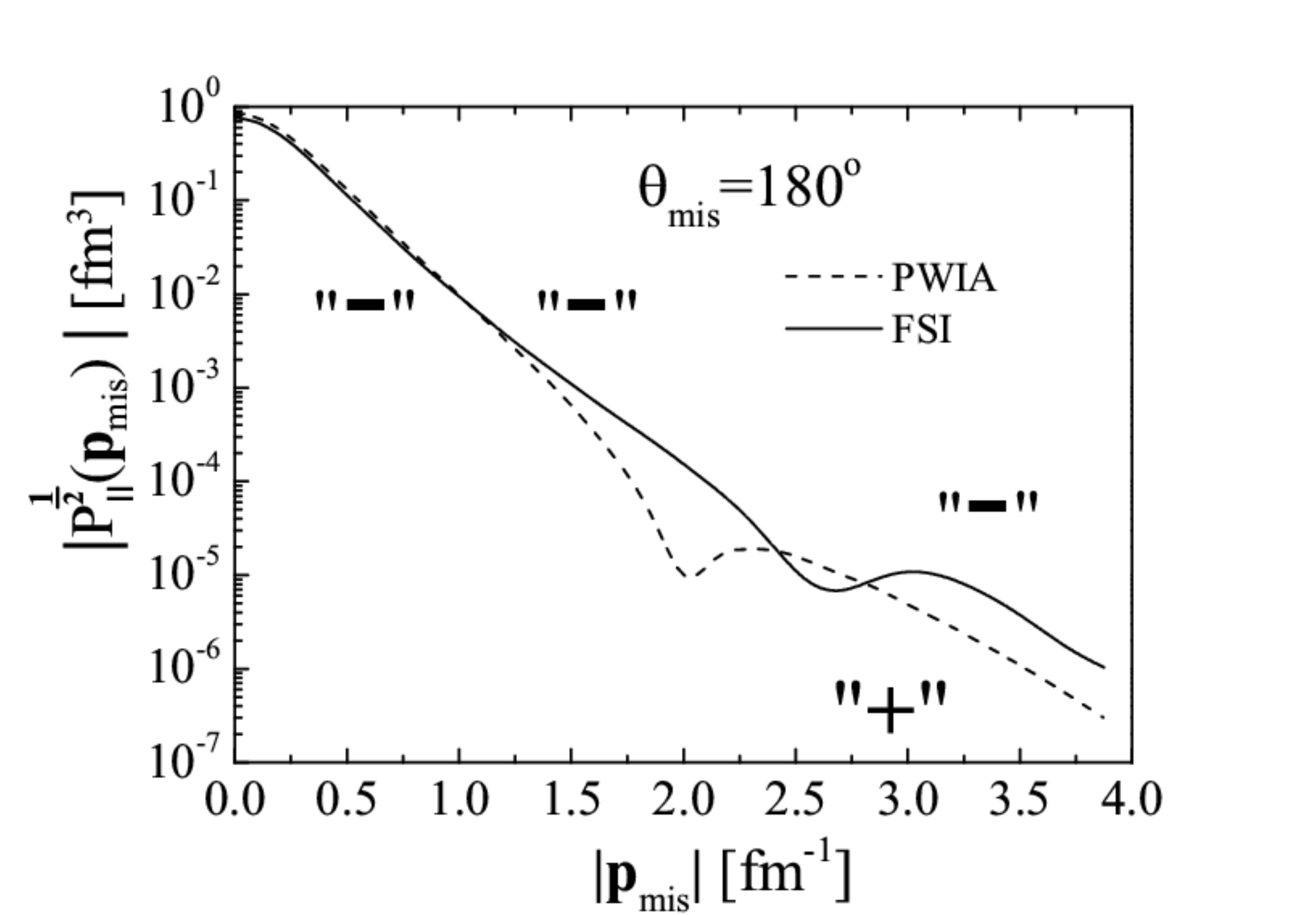}}
$~$\parbox{8cm}{\vspace{0.5cm} \includegraphics[width=8cm]{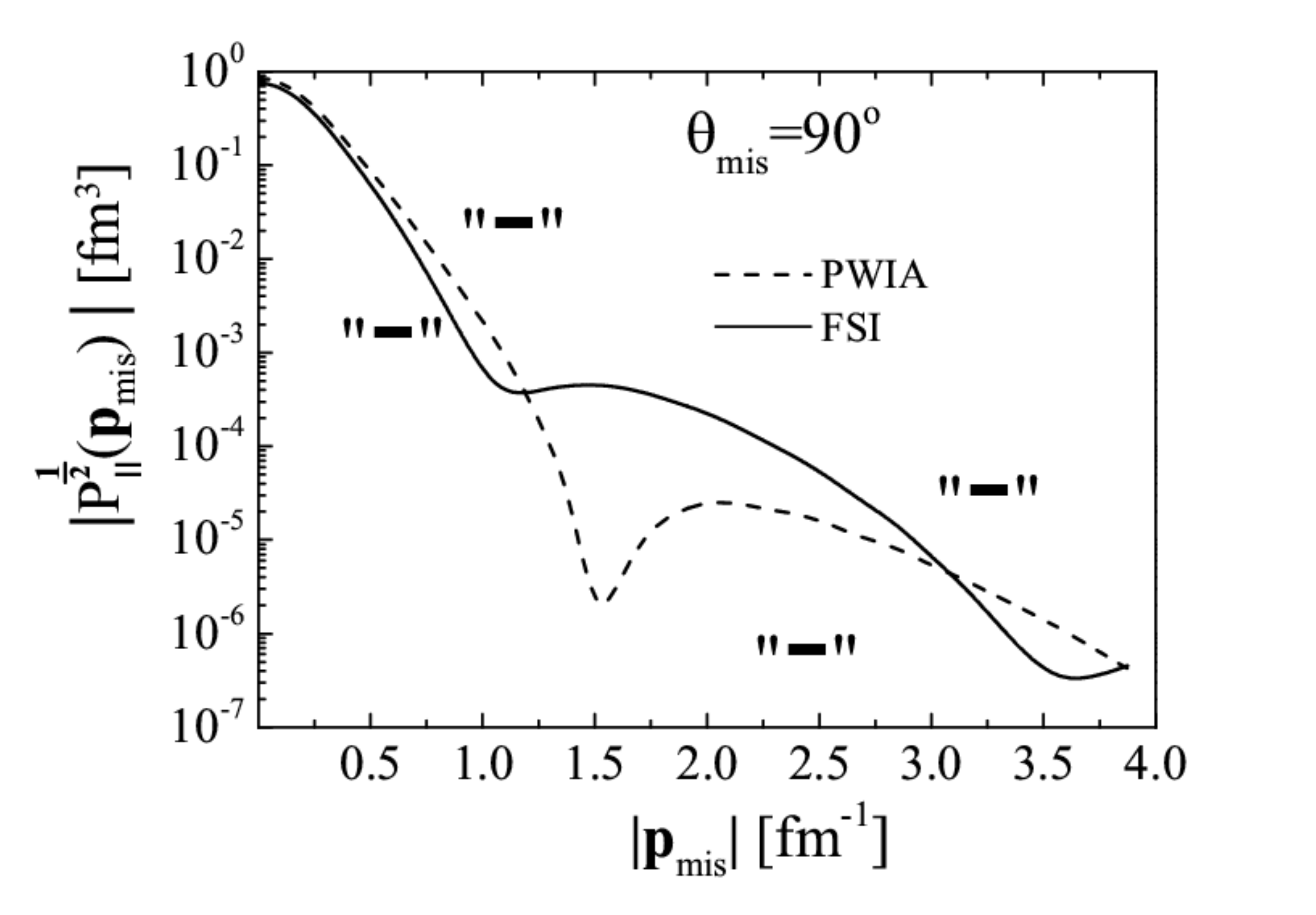}}
\caption{
{The absolute value { $\left|{\cal P}_{||}^{\frac12}\right|$}, { relevant
for a spectator SiDIS with a deuteron in the
final state}, for the
reaction $^3\vece{{\rm He} }(\vece e,e'^{\ 2}\!H)X$, in the
Bjorken limit, vs the missing
momentum (${\bf  p}_{mis}\equiv{\bf P}_{deut}$),
in parallel, $\theta_{mis}=180^o$   (left panel),
and perpendicular, $\theta_{mis}=90^o$ and $\phi_{mis}=180^o$ (right panel)
 kinematics.
Dashed line: PWIA . Solid line:   FSI
 effects.
The sign of
{{${\cal P}_{||}^{\frac12}$ is indicated by
{\large $+$} and {\large $-$}}}.  (After Ref. \cite{our}).}}
\label{figSparPerp}
\end{figure} 
Successful applications of GEA to  unpolarized SiDIS  can be found in 
Refs. \cite{ourlast,cosyn}.
In conclusion, the embedding of the GEA overlaps, Eq. (\ref{overlaps}), 
in a Poincar\'e covariant
framework  can be achieved through a generalization of the relation shown in Eq.
(\ref{sfrhd}) \cite{todo}.
\section{Results and conclusions}
\label{concl}
The impact of the FSI treatment can be appreciated by   considering  the following
quantity, needed in the evaluation of SiDIS processes and   strictly related to
the nuclear dynamics, namely 
  ${\cal P}_{||(FSI)}^{\frac12 }({\bf  p}_N,E)$, viz
 \be
 {\cal P}_{||(FSI)}^{\frac12 }({\bf  p}_N,E)=\frac12 ~
 \left[{\cal O}^{1/2,(FSI)}_{1/2,1/2}({\bf  p}_N,E)-
 {\cal O}^{1/2,(FSI)}_{-1/2,-1/2}({\bf  p}_N,E) ~+ ~c.c.\right]
 \ee
 Indeed, the evaluation was carried out for the spectator SiDIS, where a deuteron is
detected in the final state.
Figure \ref{figSparPerp} illustrates the effects of FSI, evaluated within the GEA,
 by showing  the absolute value of
  ${\cal P}_{||}^{\frac12 }$, for the above mentioned spectator SiDIS,
 as a function of the missing momentum, ${\bf p}_{mis}\equiv {\bf P}_{deut}$, in
both the parallel
($\theta_{mis}=180^o$),
 and perpendicular ($\theta_{mis}=90^o$ and $\phi_{mis}=180^o$),
 kinematics. It is interesting to note that a wise choice of the kinematical
variables can minimize the FSI effects, as
 already shown in the unpolarized case~\cite{ourlast}. As shown in 
 Fig. \ref{figSparPerp},
FSI is negligible at low values of $|{\bf p}_{mis}|$ (in this case one
has a fast  final debris, given
${\bf p}_X \sim {\bf q}$), while
it starts to be  
sizable  for $|{\bf p}_{mis}| \geq 1 fm^{-1}$, where the  equal sign holds for the
perpendicular kinematics (cf the right panel).

To illustrate the influence of the FSI effects on the extraction procedures of the
neutron information, let us consider the
preliminary comparison, for a JLAB12 kinematics \cite{He3exp}, between the PWIA
values of  both dilution factors and nucleon polarizations
that appear in Eq. (\ref{dis}) and the corresponding  calculations with FSI effects
taken into account. As shown in Table 1,
 FSI  can sizably  modify the overlaps given in Eq. (\ref{overlaps}), that are
necessary for evaluating the
quantities in Eq. (\ref{dis}). In particular,  
 about  a $ 15-20 \%$ depolarization effect of the  nucleons in 
 $^3\vece{\rm{He}}$  is produced.
  It should be pointed that,
 even if both polarizations and dilution factors in Eq. (\ref{dis}) are affected by
the presence of FSI, fortunately  their product does not.
Therefore the extraction procedure,  can be safely applied. This seems a quite 
encouraging message, from the
 phenomenological point of view, and motivates the application of such an approach,
combined with the the Poincar\'e
 covariant treatment of the nucleus tensor, to  the
 extraction procedure of the neutron SSAs from the $^3\vece{\rm{He}}$ 
 SSAs, to be considered elsewhere \cite{todo}.

\begin{table}[t]
\caption{}
\label{tab:1}       
1) PWIA: $ p_{n}$= 0.878, $ p_{p}$= -0.023, $\theta_e=30^o$, $\theta_\pi=14 ^o$\\[4mm]
 \begin{tabular}{|ccccccccccccccc|}\hline
  $E_{beam} $, &&      $x_{Bj}$&     &   $\nu$  &       & $p_\pi$  && $f_{n}(x,z)$
&& $ p_{n}~ f_{n}$ && $f_{p}(x,z)$&& $ p_{p}~ f_{p}$\\
  GeV        &\,\,\, &        &\,\,\,&  GeV      &\,\,\,&  GeV/c         &\,\,\,&   
          &\,\,\,&     &\,\,\,&     &\,\,\,&           \\ \hline
     8.8     &&  0.21         &&    7.55      && 3.40       && 0.304         &&  
0.266      && 0.348         &&   -$8.4 10^{-3}$         \\ \hline
     8.8     &&   0.29        &&    7.15      && 3.19       && 0.286         &&  
0.251     && 0.357         &&   -$8.5 10^{-3}$  \\ \hline
     8.8     &&   0.48        &&    6.36      && 2.77       &&0.257          &&  
0.225     && 0.372         &&   -$8.9 10^{-3}$  \\ \hline
     11      &&    0.21       &&    9.68      && 4.29       &&0.302          &&  
0.265     && 0.349         &&    -$8.3 10^{-3}$  \\ \hline
     11      &&   0.29        &&    9.28      &&  4.11      &&  0.285        &&  
0.250      &&  0.357        &&    -$8.5 10^{-3}$  \\ \hline
\end{tabular} 
\vskip 2mm 2) FSI:   $ p_{n}$= 0.756, 
$ p_{p}$= -0.027,
$\langle \sigma_{eff}\rangle = 71\ mb$ \\[4mm]
\begin{tabular}{|ccccccccccccccc|}\hline
  $E_{beam} $, &&      $x_{Bj}$&     &   $\nu$  &       & $p_\pi$  && $f_{n}(x,z)$
&& $ p_{n}~ f_{n}$ && $f_{p}(x,z)$&& $ p_{p}~ f_{p}$\\
  GeV        &\,\,\, &        &\,\,\,&  GeV      &\,\,\,&  GeV/c         &\,\,\,&   
          &\,\,\,&     &\,\,\,&     &\,\,\,&           \\ \hline
     8.8     &&  0.21       &&     7.55      && 3.40           &&  0.353         && 
0.267      && 0.405         &&   -$1.1 10^{-2}$         \\ \hline
     8.8     &&  0.29       &&     7.15      && 3.19           && 0.332         && 
0.251     && 0.415        &&   -$1.1 10^{-2}$  \\ \hline
     8.8     &&  0.48       &&     6.36      && 2.77           &&0.298           && 
0.225     && 0.432        && -$1.2 10^{-2}$  \\ \hline
     11     &&   0.21       &&     9.68      && 4.29           &&  0.351         && 
0.266      && 0.405         &&    -$1.10^{-2}$ \\ \hline
     11     &&   0.29       &&     9.28      &&  4.11          &&  0.331       &&
0.250       &&  0.415      &&   -$1.1 10^{-2}$ \\ \hline
\end{tabular}
\end{table}


\begin{thebibliography}{3}

\bibitem{He3exp} Cates G. et al, E12-09-018, JLAB approved experiment,
hallaweb.jlab.org\slash collab\slash PAC\slash PAC38/\slash E12-09-018-SIDIS.pdf;
X. Jiang et al, www.jlab.org\slash exp\underline{~}prog\slash generated\slash
apphalla.html;
 J. Arrington et al, 
www.jlab.org\slash $\tilde{~}$jinhuang\slash 12GeV\slash 12GeVLongitudinalHe3.pdf.

\bibitem{todo}
A. Del Dotto, L.P. Kaptari, E. Pace, G. Salm\`e and  S. Scopetta, to be published.


\bibitem{Sco}
S. Scopetta: Neutron single spin asymmetries from semi-inclusive deep 
inelastic scattering off transversely polarized $^3$He. Phys. Rev. {\bf D 75},
054005 (2007)  



\bibitem{Dotto}
A. Del Dotto, E. Pace, G. Salm\`e and S. Scopetta:
Transversity studies with a polarized 3He target.
 Il Nuovo Cimento {\bf C 35}, 101 
(2012). 

\bibitem{Pac}
E. Pace, G. Salm\`e, S. Scopetta, A. Del Dotto and  M. Rinaldi:
Neutron Transverse-Momentum Distributions and Polarized
$^3$He within Light-Front Hamiltonian Dynamics.
 Few-Body Sys. 
{\bf 54}, 1079 (2013). 

\bibitem{our} L.P. Kaptari, A. Del Dotto, E. Pace, G. Salm\`e and S.
 Scopetta:
Distorted spin-dependent spectral function of an A=3 nucleus and SiDIS processes. 
  arXiv:1307.2848.
\bibitem{ourlast}
  C.~Ciofi degli Atti and L.~P.~Kaptari:
  SiDIS off Few-Nucleon Systems: 
  Tagging the EMC Effect and Hadronization Mechanisms with Detection 
  of Slow Recoiling Nuclei. 
  Phys. Rev. {\bf C 83}, 044602 (2011). 

\bibitem{Ciofi1} C. Ciofi degli Atti, E. Pace and   G. Salm\`e:
Spin dependent spectral function of $^3$He and the asymmetry in the process 
$^3\vece{{\rm He}} (\vece e, e')X$.
Phys. Rev. {\bf C 46}, 1591 (1992).
\bibitem{Ciofi2}
C. Ciofi degli Atti, E. Pace, G. Salm\`e and  S. Scopetta:
Nuclear effects in deep inelastic scattering of polarized electrons off
polarized $^3$He and the neutron spin structure functions. 
Phys. Rev. {\bf C 48}, R968 (1993).
\bibitem{ckk}
  C.~Ciofi degli Atti, L.~P.~Kaptari and B.~Z.~Kopeliovich.
  Final state interaction effects in semiinclusive DIS off the deuteron.
  Eur.\ Phys.\ J.\ A {\bf 19}, 145 (2004).
\bibitem{cosyn}
 W.~Cosyn and M.~Sargsian. Final-state interactions in
semi-inclusive deep inelastic
scattering off the Deuteron. Phys. Rev.C {\bf 84}, 014601 (2011)

\end{thebibliography}
\end{document}